\documentstyle[11pt,newpasp,twoside,epsf]{article}
\def\edcomment#1{\iffalse\marginpar{\raggedright\sl#1\/}\else\relax\fi}
\reversemarginpar

\begin{document}
\title{The evolutionary status of The Frosty Leo Nebula}
\author{Tyler Bourke$^1$, A. R. Hyland$^2$, Garry Robinson$^3$, \& Kevin
Luhman$^1$}
\affil{$^1$Harvard-Smithsonian Center for Astrophysics,
Cambridge MA, USA\\
$^2$Southern Cross University, Lismore NSW, Australia\\
$^3$University College, UNSW-ADFA, Canberra ACT, Australia
}


\begin{abstract}
We present new observational data for IRAS 09371+1212, the Frosty Leo
Nebula, in the form of infrared spectra from 2.2--2.5\mbox{$\mu$m} which 
reveal photospheric bands of CO.  The $^{12}$CO/$^{13}$CO band ratio 
determined is similar 
to those exhibited by evolved K giant stars, and supports the proposal that 
the object is highly evolved.  The equivalent width of the CO bands implies, 
however, that the spectral type lies in the range G5III to K0III,
somewhat earlier than K7III, as derived from colours and optical spectra.  
The smaller equivalent widths of the CO bands may reflect a low metal 
abundance which could fit
well with the considerable height of the source above the galactic 
plane ($>$0.9kpc).
\end{abstract}

\section{Introduction}

The Frosty Leo Nebula, IRAS 09371+1212, remains unique with its extremely deep 
3.1\mbox{$\mu$m} absorption feature, almost two decades in depth, its twin 
far-infrared emission features at 44 and 62\mbox{$\mu$m}, and its abnormal 
location, $>$0.9kpc from the plane of the Galaxy.  Forveille et 
al.\ (1987; hereafter FMOL) concluded, primarily on the basis of its 2.6mm 
CO emission spectrum, that it is a post-AGB star 
with a bipolar circumstellar envelope, and suggested water ice mantles around 
the grains as the source of the excess emission in the 60\mbox{$\mu$m} 
IRAS band.

As noted by FMOL and Robinson, 
Smith \& Hyland (1992; hereafter RSH) the object appears to have evolved 
rapidly as its luminosity 
[250($d$/kpc)$^2$\mbox{$L_{\odot}$}]
has apparently decreased by about an order of magnitude since the 
circumstellar envelope was ejected, and the lack of significant amounts of 
warm dust indicates that dust is not currently forming near the central 
source.  RSH suggested the object could possibly be a weak-line 
T~Tauri star in the final stages of ejecting its embryonic dust shell.  

A seminal observational test, suggested by RSH, to 
determine the evolutionary status of IRAS 09371+1212, 
is to determine the $^{12}$C/$^{13}$C ratio 
from the first 
overtone photospheric bands of CO which should be present at around 
2.3\mbox{$\mu$m} 
in a late type source.  In highly evolved objects this ratio lies in the 
range of $\sim$5--20 (e.g., 
Lambert \& Ries 1981), 
while in pre-main-sequence objects and local molecular clouds the ratio 
is $>$60 (Langer \& Penzias 1993) and closer to the solar value ($\sim$90).

Here we report spectroscopic measurements of the first 
overtone bands of $^{12}$CO and $^{13}$CO at $\sim$2.3\mbox{$\mu$m} of 
IRAS 09371+1212 which have been used to determine the $^{12}$C/$^{13}$C 
isotope ratio, and so provide clear confirmation of its evolved status. 
The observations were obtained with IRIS on the AAT 
(16 Feb.\ 1992; Fig.\ 1), and Fspec on the 
Steward 2.3m Bok Reflector (19 Nov.\ 1996; Fig.\ 2).

\section{Results and Discussion}
%
%
\subsection{The $^{12}$CO/$^{13}$CO band strengths and the evolutionary status}

The abundance ratio $^{12}$C/$^{13}$C derived from the 2.3\mbox{$\mu$m}
spectrum (Fig.\ 1) is $\sim$10 (e.g., Frogel 1971).  This result 
establishes that IRAS 09371+1212 is 
an evolved object (which is likely to be on or beyond the AGB).
In support of this argument, Barnbaum, Morris \& Forveille (1996), from 
observations of $^{12}$CO and $^{13}$CO J=1-0 near 2.6mm, 
have determined that $^{12}$C/$^{13}$C may be as low as 2.  

\begin{figure}[t]
\plotfiddle{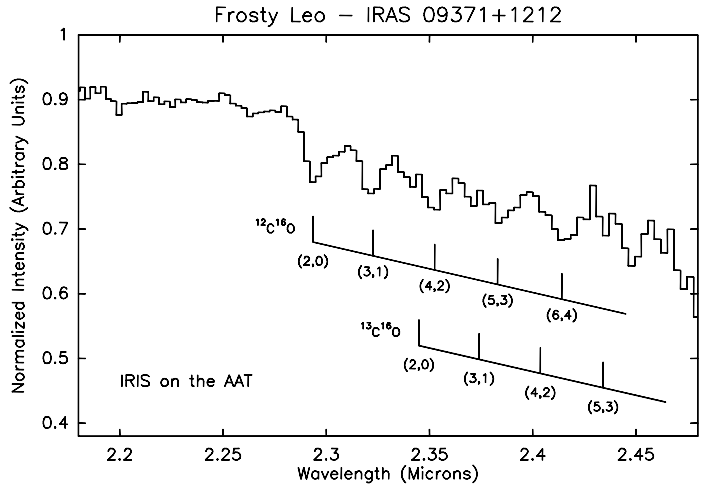}{6cm}{0}{70}{70}{-150}{0}
\caption{IRIS spectrum of IRAS 09371+1212 (resolution $\sim$450).  The
first overtone photospheric bands of $^{12}$CO are clearly visible, and
the $v=2-0$ bandhead of $^{13}$CO at 2.35\mbox{$\mu$m} is also evident.}
\end{figure}

\begin{figure}[t]
\plotfiddle{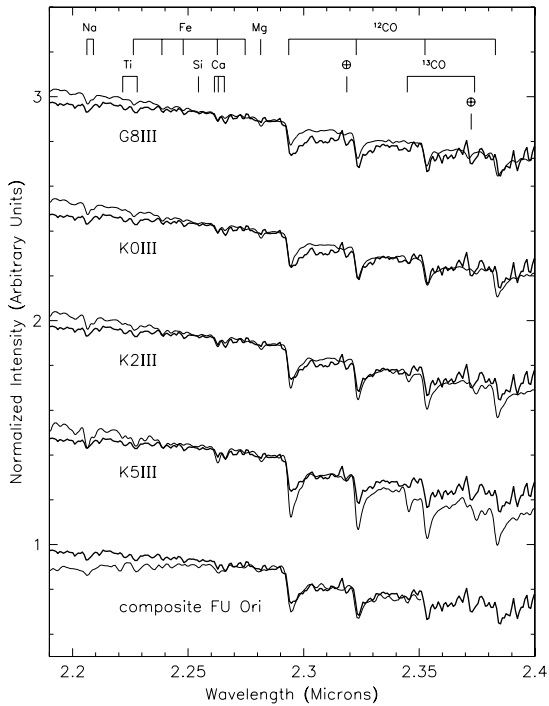}{9.1cm}{0}{38}{38}{-110}{-20}
\caption{Fspec spectrum of IRAS 09371+1212 (heavy line), with a
resolution of $\sim$1200, compared with
standard spectra and a composite FU Orionis spectrum.}
\end{figure}

%
%
\subsection{The spectral type}

The best estimate of the spectral type, based on optical spectra, is
K7II to III, with a distance of 1.0-4.3 kpc (Mauron, Le Brogne \& 
Picquette 1989).
In Figure 2 we show a high resolution spectrum of IRAS 09371+1212, along 
with spectra of giants from
Kleinmann \& Hall (1986) and a composite FU Orionis spectrum.  From this figure 
we see that both the $^{12}$CO and $^{13}$CO ($v$=2-0) bandheads match well 
with those of the K0III standard,
and the strength of the photospheric CO bands are weaker than that 
expected for a K7 giant.  The equivalent width of the CO lines is found from 
Figure 1 to be $\sim$53\AA\ which implies a J--L intrinsic colour of 
$\sim$ 0.6 (Hyland 1974), and a spectral type of 
G5 to G8, or K0 at the latest (Frogel 1971; Bessell \& Brett 1988).

Qualitatively the spectrum of IRAS 09371+1212 matches the spectrum of a
K0III star far better than that of a K5III or later star, although
none of the spectra presented in Figure 2 match
the spectrum of IRAS 09371+1212 perfectly.  
In order to match the 2.2--2.5\mbox{$\mu$m} spectrum of IRAS 09371+1212 with 
the later spectral types requires that IRAS 09371+1212
also contains a featureless continuum, thereby diluting the observed
bands.  However, the spherical circumstellar dust model of RSH
implies that the maximum grain temperature is $\sim$70K, too low to give 
rise to sufficient emission in the near infrared to 
reduce the observed strengths of these bands.  
The object therefore appears 
anomalous in regard to its spectral type. 

As seen in Figure 2, the 2\mbox{$\mu$m} spectrum of IRAS 09371+1212 is 
also well matched by the composite FU Orionis spectrum.  
However, shortward of 2.2\mbox{$\mu$m}
FU Ori stars show strong absorption due to photospheric water 
(e.g., Mould et al.\ 1978),
which is not seen in the full K-band spectrum of IRAS 09371+1212
and effectively rules out IRAS 09371+1212 being an FU Ori object.

%
%
\subsection{Consequences of the evolutionary status and the low luminosity}

The evolved nature of IRAS 0937+1212 raises a number of issues which 
need to be resolved.  At this stage of evolution the source should have 
undergone a helium flash.  However the 
luminosity is only 250($d$/kpc)$^2$\mbox{$L_{\odot}$}, and appears to 
be too low for a 
helium flash to have occurred (Lattanzio 1986).  Furthermore, if the two stars 
in the system have approximately equal magnitudes at both J and H 
(Roddier et al. 1995), then they each have a luminosity of only 
125 \mbox{$L_{\odot}$}.  This demands that either (i) the luminosity was much 
larger 
in the past, and has somehow decreased dramatically, or (ii) the object is 
much further away than current estimates.

Each of these alternatives raises further interesting issues.  Although solar 
mass stars undergo significant luminosity changes during the helium flash 
phase (Vassiliadis \& Wood 1993), theoretically these are only of the order 
of a factor of three, and it would appear most unlikely that the luminosity 
could decrease to the level of 125--250\mbox{$L_{\odot}$} from the typical 
helium flash luminosity required for a G8--K0 star undergoing thermal pulses 
of $\sim$1500\mbox{$L_{\odot}$}.  Yet, at the extreme end of the models of 
Vassiliadis \& Wood it 
can be seen that a one solar mass star undergoing a helium flash can exhibit 
a drop of a factor of seven in luminosity over a very short time scale, on 
the order of 50--100 years.  It is therefore possible that IRAS 09371+1212 
is in such a very shortlived phase of evolution.

Nevertheless, the lowest luminosity attained $\sim$700\mbox{$L_{\odot}$} 
would require that the source with its high galactic latitude of 
$b$=42\hbox{$.\!\!^\circ$}7 would be at a 
distance of 2.4 kpc, and be 1.6 kpc above the galactic plane (assuming
L$_{\star} \sim 125L_{\odot}$). 
Such a height 
might suggest that the star is either an outer disk object or has halo 
characteristics.  In either case one would expect that the source might 
exhibit a lower than solar metal abundance.  We note that the measured strength 
of the CO bands is considerably weaker than would be expected for a K7 star.  
This may be an indication of a lower CO abundance related to an overall 
metal weakness for the object, which would be consistent with the above 
views.  

Alternatively, assuming the luminosity of IRAS 09371+1212 is similar to normal 
planetary
nebula central stars ($\sim$3000\mbox{$L_{\odot}$}), its distance would then 
be 3.5kpc, and the height above the plane 2.3kpc, which is not improbable 
(Phillips \& Cuesta 1997).
It may simply be that IRAS 09371+1212 is such an object.  

%
%


%
%


\begin{references}
\reference Barnbaum, C., Morris, M., \& Forveille, T. 1996, BAAS, 189, 
number 97.15
\reference Bessell, M.S., \& Brett, J.M. 1988, PASP, 100, 1134
\reference Forveille, T., Morris, M., Omont, A., \& Likkel, L. 1987, A\&A, 
176, L13 (FMOL)
\reference Frogel, J.A. 1971, PhD thesis, California Institute of Technology
\reference Hyland, A.R. 1974, in Highlights of Astronomy, 
Vol 3, ed.\ G.\ Contopoulos (Dordrecht: Reidel) 307
\reference Kleinmann, S.G., \& Hall, D.N.B. 1986, ApJS, 62, 501
\reference Lambert, D.L., \& Ries, L.M., 1981, ApJ, 248, 228
\reference Langer, W.D., \& Penzias, A.A. 1993, ApJ, 409, 539L
\reference Lattanzio, J.C. 1986, ApJ, 311, 708
\reference Mauron, N., Le Borgne, J.-F., \& Picquette, M. 1989, A\&A, 218, 213
\reference Mould, J.R., Hall, D.N.B., Ridgway, S.T., Hintzen, P., \&
Aaronson, M. 1978, ApJ, 222, L123
\reference Phillips, J.P., \& Cuesta, L. 1997, A\&A, 326, 831
\reference Robinson, G., Smith, R.G., \& Hyland, A.R. 1992, MNRAS, 256, 437
(RSH)
\reference Roddier, F., Roddier, C., Graves, J.E., \& Northcott, M.J. 1995, 
ApJ, 443, 249
\reference Vassiliadis, E., \& Wood, P.R. 1993 ApJ 413 641
\end{references}
\end{document}